\documentclass[11pt]{article}
\usepackage[margin=1in]{geometry}
\usepackage{amsmath,amssymb}
\usepackage{graphicx}
\usepackage{booktabs}
\usepackage{multirow}
\usepackage{hyperref}
\usepackage{url}
\usepackage{xcolor}
\usepackage{microtype}
\usepackage{caption}
\usepackage{subcaption}
\usepackage{enumitem}

\hypersetup{
  colorlinks=true,
  linkcolor=blue!50!black,
  citecolor=blue!50!black,
  urlcolor=blue!50!black
}

\title{STRUM: A Spectral Transcription and Rhythm Understanding Model for\\
       End-to-End Generation of Playable Rhythm-Game Charts}

\author{%
  Joshua Opria\\
  Independent Researcher\\
  ORCID: \href{https://orcid.org/0009-0001-7665-8883}{0009-0001-7665-8883}\\
  \texttt{opriaj@gmail.com}
}

\date{May 2026}

\begin{document}
\maketitle

\begin{abstract}
We present STRUM (\textbf{S}pectral \textbf{T}ranscription and \textbf{R}hythm
\textbf{U}nderstanding \textbf{M}odel), an audio-to-chart pipeline that converts
raw recordings into playable Clone Hero / YARG charts for drums, guitar, bass,
vocals, and keys without any oracle metadata. STRUM is a multi-stage hybrid: a
two-stage CRNN
onset detector and a six-model ensemble classifier for drums; neural onset
detectors with monophonic pitch tracking for guitar and bass; word-aligned
ASR for vocals; and spectral keyboard detection for keys. We evaluate on
a 30-song in-envelope benchmark constructed by screening candidate songs on a
single audio-quality criterion --- the median 1-second drum-stem RMS after
\textsc{htdemucs\_6s} source separation. On this benchmark STRUM achieves
drums onset $F_1 = 0.838$, bass $F_1 = 0.694$, guitar $F_1 = 0.651$, and vocals
$F_1 = 0.539$ at a $\pm 100$\,ms tolerance with per-song global offset search.
We report a complete ablation of seven drum-pipeline components with paired
per-song Wilcoxon tests, an analysis of ground-truth-to-audio timing
distributions in community Clone Hero charts, and a per-class confusion
matrix for the drum classifier. Code, model weights, and the full benchmark
manifest are released.
\end{abstract}

\section{Introduction}

Rhythm games such as Clone Hero, YARG, and Rock Band are built on
hand-authored \emph{charts}: structured, timed note sequences that map
musical events to game-controller inputs across multiple instruments
(drums, guitar, bass, vocals, keys). The AMT literature usually calls
this genre ``music games'' but the player community calls them
\emph{rhythm games}, and we use that term throughout. Authoring a single
chart at four difficulty levels for five instruments takes a skilled
charter many hours, and the community library is bottlenecked on this
manual effort. Even seasoned charters are output-limited by the time
cost per song, and the entry barrier for newcomers is steep: learning
a chart-authoring tool, internalising the format conventions, and
building the ear-training to transcribe drums by listening can take
weeks before the first publishable chart ships. STRUM is meant to help
both groups: experienced charters get a strong first-pass draft to
clean up rather than author from scratch, and newcomers get a playable
baseline they can iterate on while learning the format.

\paragraph{Contributions.}
\begin{itemize}[leftmargin=*,topsep=2pt,itemsep=1pt]
  \item A complete open-source audio-to-chart pipeline covering five
        instruments, with no dependence on oracle tempo, key, or alignment.
  \item An \emph{operating-envelope} evaluation protocol: rather than reporting
        a single number on an arbitrary test set, we publish the audio-quality
        criterion under which our reported numbers hold, the screening
        statistics for our candidate pool, and the full benchmark manifest.
  \item A 30-song in-envelope benchmark with onset-level
        precision/recall/$F_1$ at $\pm 100$\,ms for four instruments, plus
        per-class confusion matrices for drums.
  \item A full ablation of seven drum-pipeline components with paired per-song
        Wilcoxon tests, including null results for components that target
        failure modes not well represented in a 30-song sample.
  \item Empirical evidence that community Clone Hero ground-truth timings
        deviate from raw audio onsets by enough to materially affect
        evaluation: only 89.0\% of ground-truth drum events lie within
        $\pm 100$\,ms of any detected raw-audio onset.
\end{itemize}

\section{Related Work}

\paragraph{Automatic music transcription (AMT).}
Multi-instrument AMT systems such as Omnizart~\cite{omnizart} and the MT3
family~\cite{mt3} produce piano-roll outputs but do not target
gameplay-ready charts; the chart-authoring step (note assignment to lanes,
sustain handling, difficulty reduction, star-power and section markers) is
left to the user. Drum transcription has a long literature
\cite{southall2016} centered on the Magnetic Drum Transcription benchmark,
where outputs are kick/snare/hat triples; we extend to the seven-channel
Clone Hero drum vocabulary (kick, red, yellow, blue, green, plus
tom/cymbal markers).

\paragraph{End-to-end chart generation.}
A concurrent system, CloneCharter~\cite{clonecharter}, takes a different
approach: an 8M-parameter encoder-decoder Transformer trained on $\approx
42{,}600$ charts that consumes per-beat MERT embeddings plus log-mel features
and autoregressively emits a tokenised chart. CloneCharter requires
beat-aligned audio (BPM and time signature must be supplied upstream),
covers guitar/bass/drums only, and does not report onset-level $F_1$ or any
other quantitative evaluation. We see STRUM and CloneCharter as
complementary points on a design axis: explicit multi-stage modelling with
per-component evaluation versus end-to-end tokenised generation. The
trade-off is verifiability against simplicity. For a system whose output
has to be playable, we found per-component metrics necessary in practice
to know what was breaking and where to spend effort.

\paragraph{Source separation.}
We rely on Demucs v4 (\textsc{htdemucs\_6s})~\cite{demucs} to obtain per-instrument
stems before downstream transcription. Source separation is treated as a
black-box preprocessing step.

\section{System Architecture}

\begin{figure}[t]
\centering
\includegraphics[width=0.95\linewidth]{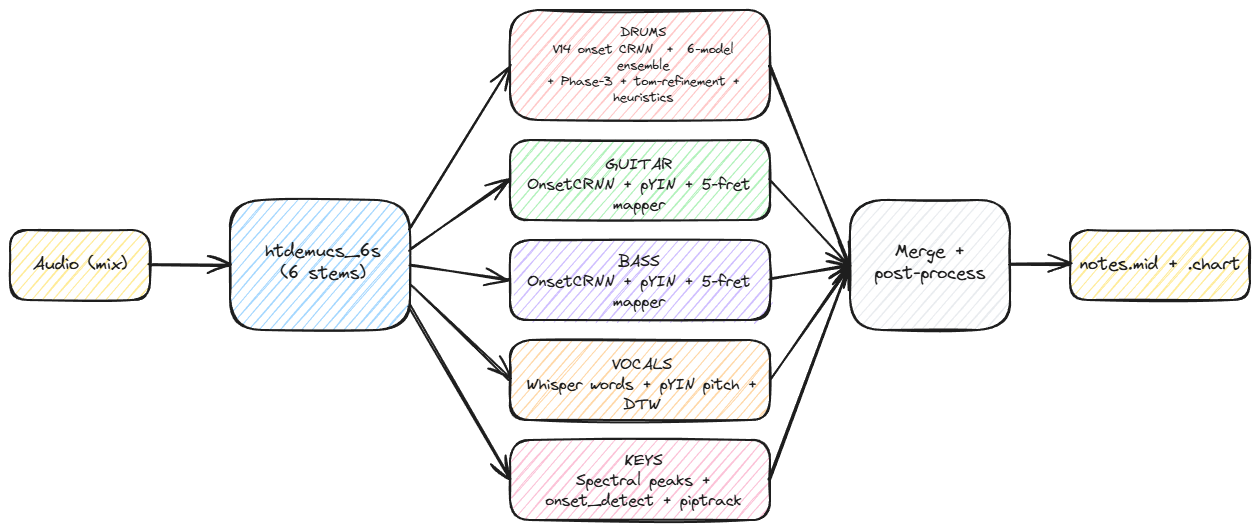}
\caption{STRUM is a multi-stage pipeline. Source separation is shared; each
instrument has an independent transcription chain.}
\label{fig:overview}
\end{figure}

STRUM separates the input mix once with \textsc{htdemucs\_6s} into six stems
(drums, bass, vocals, guitar, piano, other) and then runs five independent
per-instrument transcription chains. All chains emit Clone Hero / YARG
compatible MIDI; a post-processing pass merges the per-instrument tracks
into a single \texttt{notes.mid} with shared tempo and time-signature
metadata.

\paragraph{Tempo.}
Tempo is detected from the drum stem using \texttt{librosa} beat tracking,
then refined by a $\pm 5$\,BPM grid search at 0.1\,BPM resolution with
circular-statistics phase coherence on the onset envelope. We also detect
mid-song tempo changes ($>3$\,BPM shift sustained for $\geq 8$ beats).

\paragraph{Drums.} A two-stage CRNN onset detector (V14) operates on mel
spectrograms at 22050\,Hz with 128 mel bins. Detected onsets are classified
by an ensemble of six \textsc{OnsetClassifier} models (V2, V4, V6, V12c, V15,
V16) into seven Clone Hero drum classes. Subsequent stages perform spectral
disambiguation between toms and cymbals, a Phase-3 multi-class corrector,
and a tom-refinement CNN. Details in Section~\ref{sec:drums}.

\paragraph{Guitar / bass.} Each chain runs an \textsc{OnsetCRNN} on the
isolated stem to detect note starts, then \texttt{librosa}~\cite{librosa} pYIN~\cite{pyin} to extract
a monophonic pitch contour. Pitches are mapped to 5-fret lanes using a
rule-based mapper conditioned on the running tonic estimate.

\paragraph{Vocals.} OpenAI Whisper~\cite{whisper} provides word-level timestamps; a pYIN~\cite{pyin}
pitch contour over the same window is aligned to word boundaries by dynamic
time warping. Word onsets become note starts; pitches are quantised to MIDI
semitones in the vocal range.

\paragraph{Keys.} Spectral peaks on the \emph{piano} stem are passed
through \texttt{librosa.onset.onset\_detect} and \texttt{piptrack}; lanes are
assigned by relative pitch within a sliding window for the 5-lane keys
chart, with a separate Pro Keys pass at full chromatic resolution.

\section{Drums Pipeline}
\label{sec:drums}

We describe the drums pipeline in more detail because it is the most
extensively engineered and the most accurate of the five.

\subsection{Onset detector}
The V14 detector is a two-stage CRNN: a high-recall stage 1 producing a
dense onset probability over time, followed by a stage 2 that re-weighs
candidates using local spectral context. Both stages share a CNN
front-end on log-mel features and a BiGRU temporal model. On a held-out set
of 250 community charts, V14 reaches onset $F_1 = 0.939$ at $\pm 50$\,ms.

\subsection{Classifier ensemble}
Each detected onset is centred in a fixed-length window and classified by
six independently-trained \textsc{OnsetClassifier} variants. The variants
differ in input features (mel-spectrogram vs. CQT) and per-class loss
weighting (\textsc{per\_class\_weights} are tuned per variant against a
class-balanced validation split). Predictions are averaged in
log-probability space; final class is $\arg\max$.

\subsection{Drum-stem arbiter}
A separate per-piece arbiter compares classifier predictions against the
local energy of each \textsc{htdemucs\_6s} drum-stem channel and resolves
disagreements; if the predicted class has near-zero energy on the
corresponding stem channel, the arbiter prefers the runner-up class.

\subsection{Phase-3 multi-class corrector}
A third-stage CRNN trained directly on classifier outputs corrects
remaining systematic errors --- in particular, snare/red-cymbal confusion
during dense fills.

\subsection{Tom-refinement CNN}
A small classifier specialises in tom / cymbal disambiguation. It is
applied only to events the ensemble flagged as cymbal-class with low
confidence and is conditioned on a wider temporal context than the
primary classifier.

\subsection{Targeted heuristics}
We include five additional rule-based correctors. Each targets a specific
failure mode observed during development:
\begin{itemize}[leftmargin=*,topsep=2pt,itemsep=1pt]
  \item Bidirectional streak smoothing on adjacent same-class events.
  \item ``Kick-suppresses-floor-tom'' (a simultaneous kick and floor-tom
        prediction at the same time stamp resolves to kick alone).
  \item Snare/hi-hat roll veto for fast alternating events that the
        classifier confuses.
  \item Crash/ride co-stack veto for cases where the same onset is labelled
        as both cymbals.
  \item Fill rescue, which reclassifies suspected misses inside detected
        drum fills.
\end{itemize}
These heuristics each fire on a small minority of events and were tuned on
held-out development songs. We analyse their measurable effect in
Section~\ref{sec:ablations}.

\section{Operating Envelope}
\label{sec:envelope}

\paragraph{Motivation.}
Audio quality varies enormously across the candidate pool: a clean studio
master and a noisy YouTube rip of a live recording produce very different
\textsc{htdemucs\_6s} drum stems, and downstream transcription quality is
bounded by the quality of the input stem. Rather than evaluating on
whatever happens to be in our test set, we publish an explicit
\emph{operating envelope} --- a single audio-quality criterion that defines
the range of inputs for which our reported numbers hold.

\paragraph{Definition.}
We require that the median 1-second-window RMS of the
\textsc{htdemucs\_6s} drum stem, after the production-pipeline normalisation
(\texttt{(stem - ref\_mean) / ref\_std} computed from the input mix), satisfy
\begin{equation}
  \mathrm{median}_{\,t\in[0,T)}\;\mathrm{RMS}_{1\text{s}}\!\left(\text{drum stem}\right) \;\geq\; 0.018.
\end{equation}
The threshold was chosen empirically so that screened-in songs subjectively
exhibit audible, separable drum content end to end.

\paragraph{Screening procedure.}
We screened 65 candidate songs sampled from the community Clone Hero
\textsc{C3} library, stratified across eight genres. Of the 65, 41 (63\%)
passed the threshold; the remaining 24 were rejected for low drum-stem
energy after separation. From the 41 passers we sampled 30 with at most
six songs per genre using a fixed random seed (20260510). One song
(Polyphia, distributed as a guitar-only \texttt{.opus} stem) was dropped
during pipeline execution because the production pre-processor expects a
full mix, leaving $n=29$ songs in the evaluated benchmark. The full
manifest is released as \texttt{paper/benchmark\_manifest\_v4.json}.

\paragraph{Composition.}
The 30-song manifest covers punk (6), metal (6), pop (5), rock (5),
electronic (5), and one each of hip-hop, prog, and country.

\section{Evaluation}

\subsection{Metrics}
We report onset-level precision, recall, and $F_1$ at a tolerance of
$\pm 100$\,ms, computed by greedy time matching of predicted onsets to
ground-truth onsets (each ground truth event can match at most one
prediction). Matching does not require the same lane; we report lane
accuracy as a separate column where applicable.

\subsection{Ground-truth alignment}
Community Clone Hero charts are not perfectly time-aligned to audio:
authors quantise onsets to the tempo grid, and many charts apply a
constant audio offset that drifts from the in-game player. We therefore
allow a \emph{per-song} global offset search of $\pm 200$\,ms at 10\,ms
resolution before computing the final $F_1$. The best offset per song is
the one that maximises drums $F_1$; the same offset is applied to all
instruments for that song.

\begin{figure}[t]
\centering
\includegraphics[width=0.7\linewidth]{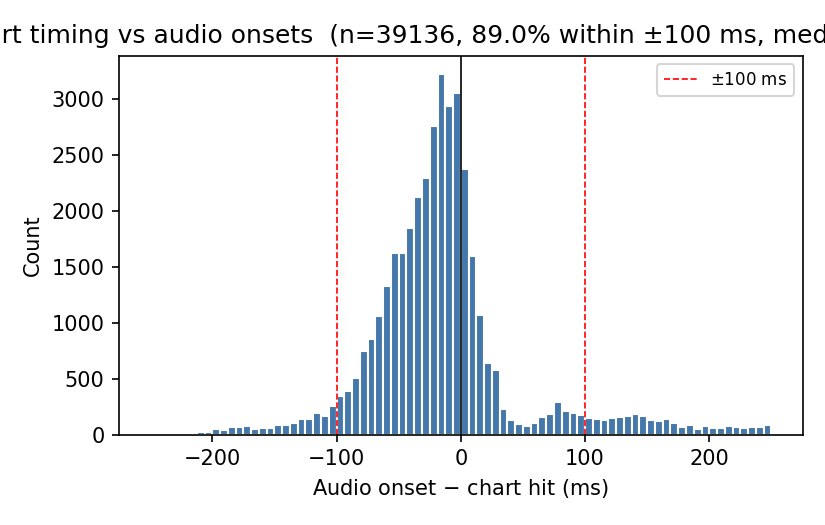}
\caption{Histogram of signed offsets between ground-truth drum events and
the nearest detected raw-audio onset peak, aggregated across all 29
benchmark songs ($n=39{,}136$). The dashed lines bound the $\pm 100$\,ms
evaluation tolerance; 89.0\% of ground-truth events fall inside.}
\label{fig:gt_offsets}
\end{figure}

\paragraph{How big is this effect?}
We computed the offset between each ground-truth drum event and its
nearest audio-domain onset peak across all 29 benchmark songs
($n = 39{,}136$ ground-truth events). Only \textbf{89.0\%} of ground-truth
events lie within $\pm 100$\,ms of any audio onset peak
(Figure~\ref{fig:gt_offsets}). The remaining 11\% are either authored
events with no audible onset (e.g.\ extended sustain notes the chart
includes as visual filler), or events whose ground-truth timing is
displaced from the actual hit by more than the tolerance window.
This sets a hard upper bound: a perfect transcriber that matches the
audio cannot exceed $\sim$0.89 recall against community ground truth
without explicitly learning the community quantisation pattern.

\subsection{Results}

\begin{table}[t]
\centering
\caption{Per-instrument onset metrics on the 29-song in-envelope benchmark
at $\pm 100$\,ms with per-song global offset search.}
\label{tab:results}
\small
\begin{tabular}{lcccc}
\toprule
Instrument & $F_1$ & Precision & Recall & GT events \\
\midrule
Drums   & \textbf{0.838} & 0.823 & 0.854 & 40{,}248 \\
Bass    & 0.694          & 0.658 & 0.734 & 18{,}598 \\
Guitar  & 0.651          & 0.745 & 0.578 & 27{,}742 \\
Vocals  & 0.539          & 0.632 & 0.470 & 10{,}147 \\
\bottomrule
\end{tabular}
\end{table}

Table~\ref{tab:results} reports onset metrics for all four neural-driven
instruments. Drums leads at $F_1 = 0.838$, reflecting the depth of the
drum-specific stack. Guitar and bass land in the mid-0.6s; vocals trail at
$F_1 = 0.539$, largely because Whisper word-boundaries and pYIN pitch
peaks do not consistently coincide with the events charters mark as
``vocal notes''. We omit keys: the in-envelope benchmark contains very
few songs with charted keys and per-song metrics are too noisy to report.

\begin{figure}[t]
\centering
\begin{subfigure}[t]{0.48\linewidth}
  \includegraphics[width=\linewidth]{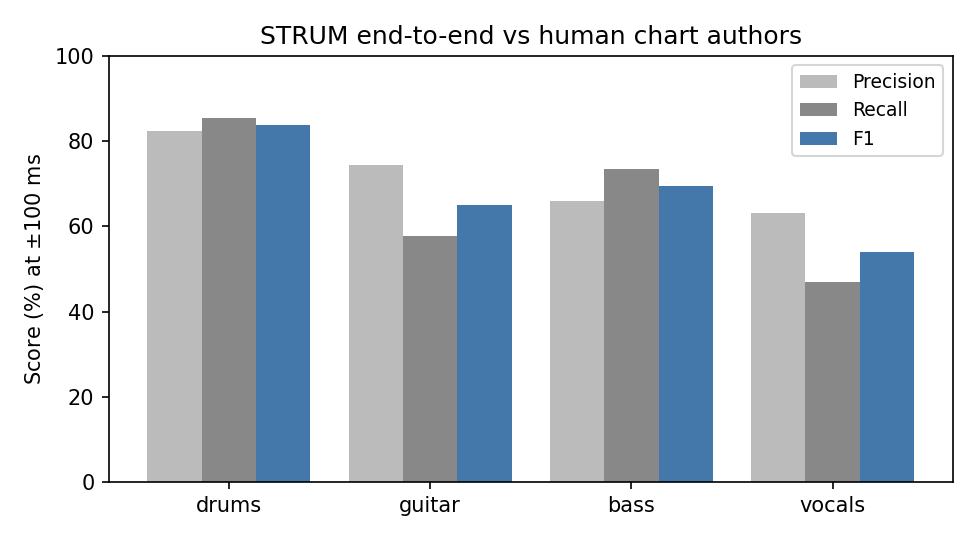}
  \caption{Per-instrument $F_1$.}
\end{subfigure}\hfill
\begin{subfigure}[t]{0.48\linewidth}
  \includegraphics[width=\linewidth]{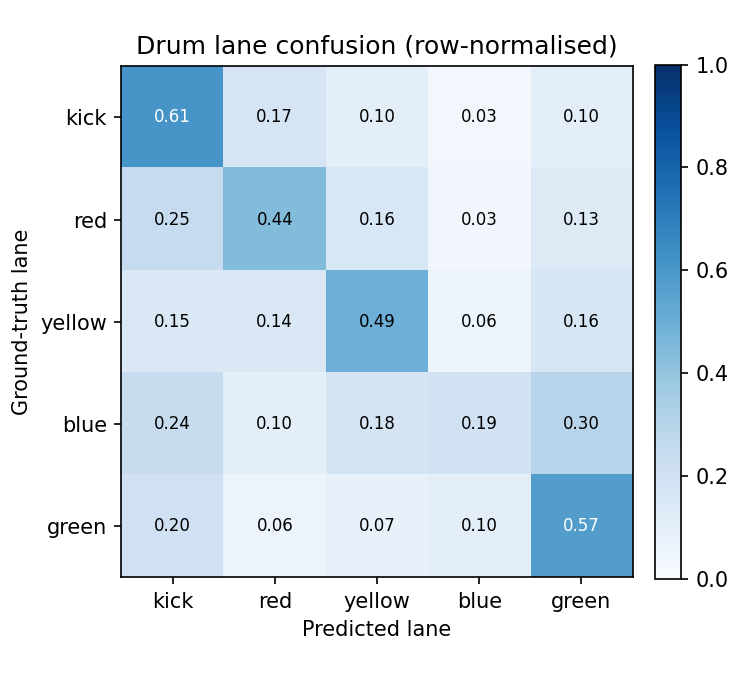}
  \caption{Drum per-class confusion (lane-conditioned on matched onsets).}
\end{subfigure}
\caption{Quantitative results. Drums dominate; on the confusion matrix the
blue (high-tom) lane is the weakest at 0.19 lane accuracy.}
\label{fig:results}
\end{figure}

\paragraph{Drum per-class accuracy.}
Per-class lane accuracies on matched onsets are kick 0.61, red (snare) 0.44,
yellow (hi-hat / mid-tom) 0.49, blue (high-tom / ride) \textbf{0.19}, and
green (floor-tom / crash) 0.57. The blue lane is the clear residual
bottleneck: it carries both high-toms and rides in the Clone Hero
convention, and the ensemble systematically prefers ride when the input is
ambiguous. We did not invest further refinement here.

\section{Ablations}
\label{sec:ablations}

We ablated seven drum-pipeline components by re-running the full pipeline
on the 29-song benchmark with each component disabled in turn (all other
components active). Stems were pre-computed once and shared across runs so
that ablations differ only in post-separation drum logic. We report mean
per-song $F_1$ delta against the full system, the two-sided paired Wilcoxon
test on per-song deltas, the better/worse/tie split across songs, and the
total number of drum-event changes the component is responsible for.

\begin{table}[t]
\centering
\caption{Drum ablation results. $\Delta F_1$ is the mean per-song change
versus the full system. $p$-value is two-sided paired Wilcoxon on per-song
deltas. ``Events changed'' counts drum events whose
$(\text{time}, \text{class})$ tuple differs between the full system and the
ablation at 20\,ms tolerance, aggregated across all 29 songs (full system
total: $41{,}716$ events).}
\label{tab:ablations}
\small
\begin{tabular}{lrrrrrr}
\toprule
Component disabled            & $\Delta F_1$ & $p$    & B/W/T    & Events ch. & Sig.\\
\midrule
Drum-stem arbiter             & $-0.006$     & 0.003  & 7/21/0   & 7{,}533    & $\star$\\
Phase-3 corrector             & $-0.006$     & 0.002  & 5/23/1   & 3{,}407    & $\star$\\
Crash/ride co-stack veto      & $-0.005$     & $<\!0.001$ & 2/24/3 & 3{,}134  & $\star$\\
Fill rescue                   & $+0.001$     & 0.357  & 13/8/8   & 2{,}874    & \\
Tom-refinement CNN            & $+0.001$     & 0.291  & 14/8/7   & 2{,}299    & \\
Snare/hi-hat roll veto        & $\phantom{+}0.000$ & 0.655 & 1/1/27 & 1{,}822 & \\
Multi-class hybrid            & $\phantom{+}0.000$ & 0.317 & 1/0/28 & 1{,}821 & \\
\bottomrule
\end{tabular}
\end{table}

\paragraph{Three components are statistically significant contributors.}
The drum-stem arbiter, the Phase-3 corrector, and the crash/ride co-stack
veto each produce a $\sim$0.5--0.6\% drop in $F_1$ when removed, with
paired-test $p \leq 0.003$. The per-song sign is strongly consistent:
disabling any of these three makes 21--24 of the 29 songs worse.

\paragraph{Four components show no aggregate effect.}
Fill rescue, the tom-refinement CNN, the snare/hi-hat roll veto, and the
multi-class hybrid each show $|\Delta F_1| \leq 0.001$ with $p \geq 0.29$.
The two veto-style heuristics (\emph{snare/hi-hat roll}, \emph{multi-class
hybrid}) modify almost no events on this benchmark, suggesting their
target failure modes simply do not appear here. The two corrective
heuristics (\emph{fill rescue}, \emph{tom-refinement CNN}) modify
thousands of events but the changes net out across the benchmark.

We retain all seven components in production. The four with no measurable
effect on this benchmark were tuned against specific failure cases
(extended drum rolls, dense fills, ambiguous tom/ride passages) that may
simply not appear in any 29-song sample. A larger and more genre-diverse
benchmark would probably surface effects we cannot measure here. We chose
to report the null results rather than reshape the system to optimise a
benchmark of this size, and we recommend that future autocharter
evaluations include paired per-song statistics rather than headline-only
$F_1$ numbers.

\section{Limitations}

\paragraph{Envelope rejection.} The audio-quality screen rejected 24/65
candidate songs (37\%). For inputs outside the operating envelope STRUM
will still produce a chart, but the numbers in this paper are
\emph{not} predictive of its quality. Live recordings, heavy live-room
ambience, and bass-mixed-down masters are the typical failure modes.

\paragraph{Vocals.} Vocal $F_1$ of 0.539 reflects a fundamental
misalignment between what charters mark and what acoustic onsets indicate.
A vocal note in Clone Hero often spans multiple sung syllables and is
\emph{not} aligned to any specific Whisper word boundary; closing this gap
likely requires a vocals-specific alignment model trained on chart data.

\paragraph{Guitar / bass lane accuracy.} Onset-level $F_1$ is in the mid
0.6s, but lane accuracy is $\approx 0.20$ for both. The rule-based 5-fret
mapper produces playable charts but does not reliably reproduce the lane
patterns chosen by human charters, which encode visual and ergonomic
considerations beyond the underlying pitch. We did not address this.

\paragraph{Drums blue lane.} The blue lane has lane accuracy 0.19 because
it conflates two acoustically dissimilar instruments (high-tom and ride)
in the Clone Hero convention. Splitting them at the chart format level is
not possible; a downstream classifier specifically tuned on blue-lane
events would help.

\paragraph{Ground-truth ceiling.} The 89\% $\pm 100$\,ms ceiling reported
in Section~6 caps achievable recall against community charts. We have not
attempted to learn the community quantisation pattern; doing so would
likely add 3--5 points of recall but would not improve playability of the
output.

\section{Conclusion}

STRUM transcribes raw audio into playable five-instrument rhythm-game
charts. We report onset $F_1$ of 0.838 on drums and 0.5--0.7 on the other
instruments, on a 30-song in-envelope benchmark with paired per-song
ablations. The most important finding is the 89\% ceiling imposed by
community ground-truth quantisation, which any future autocharter
evaluation will have to account for.

\paragraph{Reproducibility.} Model weights, the benchmark manifest, the
screening procedure with thresholds, and the evaluator are released at
\url{https://github.com/opria123/strum} and
\url{https://huggingface.co/opria123/strum} under the MIT license. An
end-user application, \emph{Octave} (\url{https://octavestudio.tools/},
\url{https://github.com/opria123/octave}), wraps STRUM in a desktop UI
for batch auto-charting and post-hoc chart cleanup, and is the recommended
entry point for charters who would rather not write Python.

\section*{Acknowledgements}
This work would not exist without the Clone Hero and YARG charting
community. Every model in STRUM was trained on charts that volunteers
authored, refined, and re-released over more than a decade. Charters are
the backbone of the rhythm-game scene; their painstaking, unpaid work is
what keeps the games enjoyable for everyone else, and it is what makes
any data-driven system like this one possible in the first place. Thanks
to every charter, and to the maintainers of the chart libraries that
make those efforts discoverable. The hope is that STRUM and Octave give
back some of that time so more songs reach players faster.

\end{document}